\documentclass[twocolumn,showpacs,aps,prb,superscriptaddress,floatfix]{revtex4}
\usepackage{graphicx}
\usepackage{dcolumn}
\usepackage{bm}
\begin{document}
\preprint{}
%

\affiliation{Department of Physics and Astronomy, McMaster University,
Hamilton, Ontario, L8S 4M1, Canada}
\affiliation{Canadian Institute for Advanced Research, 180 Dundas St. W.,
Toronto, Ontario, M5G 1Z8, Canada}

\author{J.P. Clancy}
\affiliation{Department of Physics and Astronomy, McMaster University,
Hamilton, Ontario, L8S 4M1, Canada}

\author{B.D. Gaulin}
\affiliation{Department of Physics and Astronomy, McMaster University,
Hamilton, Ontario, L8S 4M1, Canada}
\affiliation{Canadian Institute for Advanced Research, 180 Dundas St. W.,
Toronto, Ontario, M5G 1Z8, Canada}

\author{J.P. Castellan}
\affiliation{Department of Physics and Astronomy, McMaster University,
Hamilton, Ontario, L8S 4M1, Canada}

\author{K.C. Rule}
\affiliation{Department of Physics and Astronomy, McMaster University,
Hamilton, Ontario, L8S 4M1, Canada}

\author{F.C. Chou}
\affiliation{Center for Condensed Matter Sciences, National Taiwan University, Taipei 106, Taiwan.}

\title{Suppression of the commensurate spin-Peierls state in Sc-doped TiOCl}

\begin{abstract}
We have performed x-ray scattering measurements on single crystals of the doped spin-Peierls compound Ti$_{1-x}$Sc$_{x}$OCl (x = 0, 0.01, 0.03). These measurements reveal that the presence of non-magnetic dopants has a profound effect on the unconventional spin-Peierls behavior of this system, even at concentrations as low as 1 \%.  Sc-doping suppresses commensurate fluctuations in the pseudogap and incommensurate spin-Peierls phases of TiOCl, and prevents the formation of a long-range ordered spin-Peierls state.  Broad incommensurate scattering develops in the doped compounds near T$_{c2}$ $\sim$ 93 K, and persists down to base temperature ($\sim$ 7 K) with no evidence of a lock-in transition.  The width of the incommensurate dimerization peaks indicates short correlation lengths on the order of $\sim$ 12 {\AA} below T$_{c2}$.  The intensity of the incommensurate scattering is significantly reduced at higher Sc concentrations, indicating that the size of the associated lattice displacement decreases rapidly as a function of doping.

\end{abstract}

\maketitle

\section{Introduction}

Low dimensional quantum magnetic systems have been a subject of great interest, particularly since the discovery of high temperature superconductivity in the lamellar copper oxides\cite{Review}.  One such family of quantum magnets are spin-Peierls materials, quasi-one-dimensional systems which dimerize at low temperatures to form a non-magnetic singlet ground state\cite{Spin-Peierls}.  These systems typically possess chains of antiferromagnetically coupled spin 1/2 Heisenberg magnetic moments and strong magneto-elastic coupling.

Although the first materials to display a spin-Peierls transition were organic compounds, such as TTF-CuBDT\cite{Spin-Peierls} and MEM-(TCNQ)$_{2}$\cite{MEM}, studies of these materials were often limited by sample quality and a relatively low density of magnetic moments.  The discovery of the first inorganic spin-Peierls compound, CuGeO$_{3}$, with T$_{SP}$ $\sim$ 14K\cite{CuGeO3}, allowed for the growth of large, high quality single crystal samples, as well as the opportunity to introduce a wide range of magnetic and non-magnetic dopants into the system and study the resulting perturbations.  Studies of doped CuGeO$_{3}$ have been carried out wherein Cu$^{2+}$ (S=1/2) is substituted for Zn (S=0)\cite{Hase, Oseroff, Schoeffel, Martin, Grenier, Manabe, Lumsden}, Mg (S=0)\cite{Grenier, Mgdoped}, Cd (S=0)\cite{Lumsden}, Ni (S=1)\cite{Oseroff, Grenier}, Co (S=3/2)\cite{Anderson} and Mn (S=5/2)\cite{Oseroff}, or Ge$^{4+}$ (S=0) is substituted for Si (S=0)\cite{Oseroff, Schoeffel, Grenier, Lumsden}.  These experiments have revealed a rich variety of impurity induced effects, including (i) creation of free spin 1/2 moments, (ii) reduction of correlation lengths in the spin-Peierls state, (iii) depression of the spin-Peierls transition temperature, T$_{SP}$, and (iv) eventual destruction of spin-Peierls order above a critical doping, x$_{c}$.  Furthermore, the introduction of non-magnetic impurities has been shown to give rise to three-dimensional antiferromagnetic order below a second transition temperature, T$_{N}$\cite{Oseroff, Martin, Grenier, Mgdoped}, which either coexists with the dimerized ground state or replaces it.

Interest in spin-Peierls systems has recently been rekindled by the discovery of unconventional spin-Peierls behavior in quantum magnets based on Ti$^{3+}$ (3d$^{1}$).  The titanium oxyhalides TiOX (X = Cl, Br) have been shown to exhibit dimerized singlet ground states at low temperatures\cite{Shaz, Sasaki}.  However, in contrast to conventional spin-Peierls materials, TiOCl and TiOBr undergo not one, but two successive phase transitions with increasing temperature\cite{Seidel, Imai, Krimmel, Abel, Ruckamp, Hemberger, Sasaki}.  At T$_{c1}$ there is a discontinuous phase transition between commensurate and incommensurate spin-Peierls states, followed by a continous transition at T$_{c2}$ to a disordered pseudogap state.  In TiOCl, the pseudogap phase extends from T$_{c2}$ up to T$^*$ $\sim$ 130K, and throughout this range it displays an NMR signature very similar to that of the underdoped high T$_{C}$ cuprate superconductors\cite{Imai}.  TiOCl and TiOBr are also distinguished among spin-Peierls systems due to their unusually high transition temperatures (T$_{c1}$/T$_{c2}$ $\sim$ 63K/93K and 27K/47K for TiOCl\cite{Seidel} and TiOBr\cite{Sasaki} respectively) and the surprisingly large size of their singlet-triplet energy gap ($\sim$ 430 to 440K)\cite{Imai, Baker}.

TiOCl crystallizes into the orthorhombic space group {\it Pmmn} with room temperature lattice parameters of a = 3.79 {\AA}, b = 3.38 {\AA}, and c = 8.03 {\AA}\cite{Schafer}.  This structure consists of buckled Ti-O bilayers which are separated by double layers of Cl atoms and stacked vertically along the {\bf c}-axis\cite{Seidel, Schafer}.  Previous x-ray scattering results indicate that the lattice dimerization associated with the low temperature commensurate spin-Peierls state occurs along the crystallographic {\bf b}-axis\cite{Shaz}.  Between T$_{c1}$ and T$_{c2}$ this dimerized structure is also incommensurately modulated along both the {\bf a} and {\bf b} directions\cite{Krimmel, Abel}.  Measurements of the high temperature magnetic susceptibility have been described by a model of spin 1/2 Heisenberg chains, with a nearest neighbour exchange coupling of J $\sim$ 660K\cite{Seidel}.  The effective reduction of magnetic dimensionality which occurs within the Ti-O bilayers has been attributed to the ordering of Ti d$_{xy}$ electronic orbitals\cite{Seidel}. 

While TiOCl and TiOBr have attracted considerable attention, little is known about the influence of impurities on the unconventional spin-Peierls ground state of compounds such as Ti$_{1-x}$Sc$_{x}$OCl.  In this material, non-magnetic Sc$^{3+}$ (S=0) ions are substituted onto Ti$^{3+}$ (S=1/2) sites, in analogy with Zn or Mg doping in CuGeO$_{3}$ and the cuprate superconductors.  While the earliest bulk characterization of Ti$_{1-x}$Sc$_{x}$OCl provided indications of a glassy magnetic state\cite{Beynon}, more recent measurements suggest that $\chi$(T) can be modelled by finite chains of Heisenberg S=1/2 moments\cite{Seidel}.  The presence of a large Curie tail at low temperatures is consistent with the freeing of spin 1/2 moments due to the disruption of singlet pairs by impurities\cite{Beynon,Seidel}.  To date, no evidence of coexisting antiferromagnetic order has been detected down to 2K, even in samples containing up to 3 \% Sc\cite{APS}.  This result is particularly intriguing given that antiferromagnetic order has been observed in lightly doped Cu$_{1-x}$Zn$_x$GeO$_{3}$ with x as low as 0.0011\cite{Manabe}, and some theoretical models predict that long-range antiferromagnetic order should arise in disordered spin-Peierls systems for arbitrarily small dopings\cite{Fukuyama}.

In this paper we present x-ray scattering measurements on single crystal Ti$_{1-x}$Sc$_{x}$OCl which show this unconventional spin-Peierls ground state to be very sensitive to the presence of quenched, non-magnetic Sc impurities, even at concentrations on the order of only 1 \%.  Sc-doping suppresses the commensurate dimerization fluctuations observed in the incommensurate spin-Peierls and pseudogap phases of the pure material\cite{Clancy}, and inhibits the formation of a commensurate long-range ordered state down to the lowest temperatures measured ($\sim$ 7K). A short-range ordered, incommensurately-modulated state is found at all temperatures below $\sim$ T$_{c2}$. 

\section{Experimental Details}

Single crystal samples of TiOCl, Ti$_{0.99}$Sc$_{0.01}$OCl, and Ti$_{0.97}$Sc$_{0.03}$OCl were prepared using the chemical vapor transport method, as described in reference 16.  The dimensions of the pure TiOCl sample were approximately 2.0 $\times$ 2.0 $\times$ 0.1 mm, while the 1 \% and 3 \% Sc-doped samples were roughly 2.0 $\times$ 1.0 $\times$ 0.1 and 1.0 $\times$ 1.0 $\times$ 0.1 mm in size respectively.  Samples were mounted on the cold finger of a closed cycle refrigerator and aligned within a Huber four circle diffractometer.  The temperature stability of the samples was maintained at $\sim$ $\pm$ 0.01K.  X-ray scattering measurements were performed using Cu-K$\alpha$ radiation ($\lambda$ = 1.54 \AA) produced by an 18 kW rotating anode source with a vertically focused pyrolytic graphite monochromator.

\begin{figure}
\includegraphics{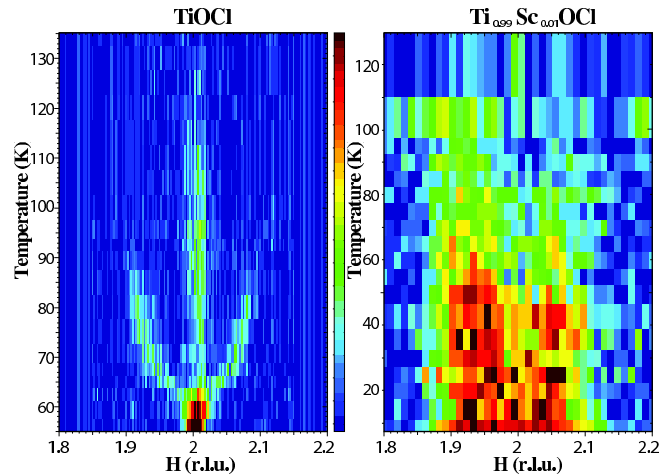}
\caption{(Color) Color contour maps of x-ray scattering as a function of temperature in Ti$_{1-x}$Sc$_{x}$OCl with x=0 (left) and x=0.01 (right).  These maps are composed of H-scans of the form (H, 1.5, 1), as shown in Fig. 2, from which a high temperature background scan has been subtracted.}
\end{figure}

X-ray scattering scans were carried out at several equivalent commensurate ordering wave vectors, (H, K+1/2, L), where superlattice Bragg peaks are expected to arise as the result of spin-Peierls dimerization.  These scans ultimately focused at the (2, 1.5, 1) position in reciprocal space, as it was here that the observed superlattice peak intensities were found to be strongest.  Scans of the form (H, 1.5, 1), (2, K, 1) and (2, 1.5, L) were performed in the H, K, and L directions of reciprocal space respectively.  A high temperature background data set, collected at 150K (Ti$_{0.99}$Sc$_{0.01}$OCl, Ti$_{0.97}$Sc$_{0.03}$OCl) or 200K (TiOCl), was then subtracted from each scan.  This was necessary due to weak $\lambda$/2 contamination of the incident beam, which results in higher order Bragg scattering at the commensurate (2, 1.5, 1) position.  This $\lambda$/2 scattering is comparable in strength to the incommensurate scattering observed in the doped samples ($\sim$ 0.1 cts/s), however, due to its lack of temperature dependence it can easily be eliminated by a simple background subtraction.

\section{Results and Discussion}

Figure 1 shows a comparison of the observed scattering in TiOCl and Ti$_{0.99}$Sc$_{0.01}$OCl.  The two colour contour maps illustrate the temperature dependence of H-scans performed through (2, 1.5, 1) in both the pure and Sc-doped samples.  It is evident that the introduction of quenched non-magnetic impurities, even at the 1 \% level, has a profound effect on the spin-Peierls behaviour of TiOCl.  In the pure compound, the growth of commensurate dimerization fluctuations can be observed throughout the pseudogap phase, from T$^*$ $\sim$ 130K to T$_{c2}$ $\sim$ 93K.  In the incommensurate spin-Peierls phase, between T$_{c2}$ and T$_{c1}$ $\sim$ 63K, these fluctuations compete and coexist with incommensurate satellite peaks which arise at (2 $\pm$ $\delta$, 1.5, 1).  At T$_{c1}$, the system locks into a commensurate long-range ordered spin-Peierls phase, which is maintained to low temperatures.  In contrast, Ti$_{0.99}$Sc$_{0.01}$OCl shows no sign of commensurate fluctuations in either the pseudogap or incommensurate spin-Peierls phases, and there is no evidence of commensurate long-range order down to 
$\sim$ 7K.  As in the pure material, incommensurate scattering is found to develop in the doped compound near T$_{c2}$ $\sim$ 93K.  However, the incommensurate scattering in Ti$_{0.99}$Sc$_{0.01}$OCl is dramatically broader than that of TiOCl and there is no lock-in transition to a commensurate state near T$_{c1}$.  Hence, a short-range ordered incommensurate phase is observed at all temperatures below T$_{c2}$. 

Representative H-scans for Ti$_{1-x}$Sc$_{x}$OCl with x=0, 0.01, and 0.03, from which the color contour map in Figure 1 was made, are shown in Figure 2.  The H-scan for TiOCl was taken at T = 75K, in the middle of the incommensurate spin-Peierls phase.  This scan clearly demonstrates the coexistence of commensurate and incommensurate scattering between T$_{c1}$ and T$_{c2}$.  The H-scans for the x=0.01 and 0.03 samples were collected at base temperature (T $\sim$ 7.5K and 7K respectively), where the dimerization scattering is fully developed.  Since the H-width of the incommensurate peaks in the doped samples is approximately equal to the magnitude of the incommensurate modulation wave vector, it is difficult to determine whether the scattering at the commensurate position is merely reduced or completely absent.  However, an upper bound can be placed on the relative intensity of the commensurate scattering, which must be more than a factor of 15 times weaker in Ti$_{0.99}$Sc$_{0.01}$OCl than TiOCl.  The K and L scans which we performed showed no significant temperature dependence in either the peak width or the peak position. The experimental resolution was sufficiently broad in the K-direction that the secondary incommensuration reported for TiOCl at (H $\pm$ $\delta$, K $\pm$ $\epsilon$, L)\cite{Krimmel, Abel} is expected to be unobservable.  Thus, our measurements are sensitive to the component of the incommensurate modulation perpendicular to the the dimerized chains ($\delta$), but not the component which lies parallel to the chain direction ($\epsilon$).

\begin{figure}
\includegraphics{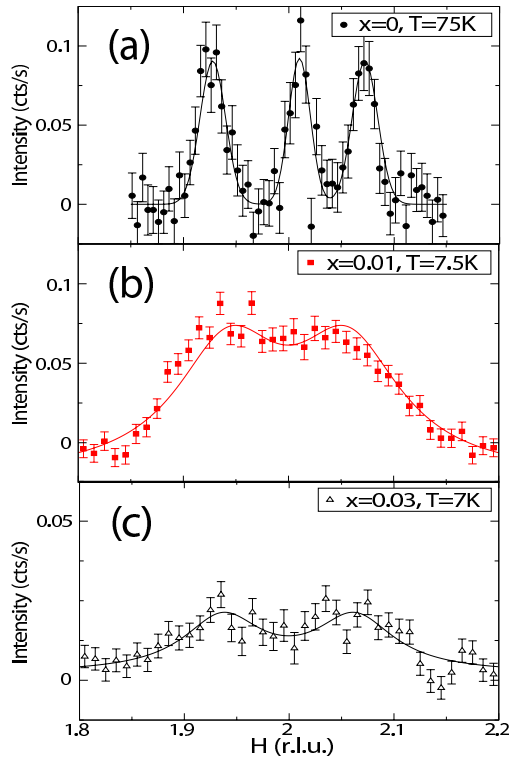}
\caption{(Color online) Representative H-scans of the form (H, 1.5, 1) in Ti$_{1-x}$Sc$_{x}$OCl.  Scans taken in the incommensurate spin-Peierls phases of: (a) TiOCl at 75 K.  (b) Ti$_{0.99}$Sc$_{0.01}$OCl at 7.5 K.  (c) Ti$_{0.97}$Sc$_{0.03}$OCl at 7 K.  The solid lines represent fits to Eqs. 1 and 2, as described in the text.}
\end{figure}

The data shown in Figure 2 was fit to appropriate forms, allowing the incommensurate ordering wavevectors and correlation lengths to be extracted as a function of temperature for Ti$_{1-x}$Sc$_{x}$OCl.  For pure TiOCl, such fits employed three Lorentzian peaks and a one-dimensional resolution convolution, suitable for near-long range ordered lineshapes as shown in Figure 2a.  For the x=0.01 and x=0.03 samples, fits were performed using a simple two peak model with Lorentzian lineshapes.  Since the widths of the incommensurate peaks in the doped samples are extremely broad, as shown in Figures 2b and 2c, no resolution convolution was necessary. The relevant 2 and 3 peak fit functions employed were: 
\begin{eqnarray}
I_{2 peak}(H) = A\left[\frac{1}{\left(\frac{H-2+\delta}{\Gamma}\right)^{2}+1}+\frac{1}{\left(\frac{H-2-\delta}{\Gamma}\right)^{2}+1}\right]
\end{eqnarray}
\begin{eqnarray}
I_{3 peak}(H) = I_{2 peak}(H) + \frac{B}{\left(\frac{H-2+\delta_{2}}{\Gamma_{2}}\right)^{2}+1}
\end{eqnarray}  
where $\delta$ is the incommensurate ordering wave vector and $\Gamma$ = a/(2$\pi$ $\xi$) is the half width at half maximum and the inverse correlation length.  

\begin{figure}
\includegraphics{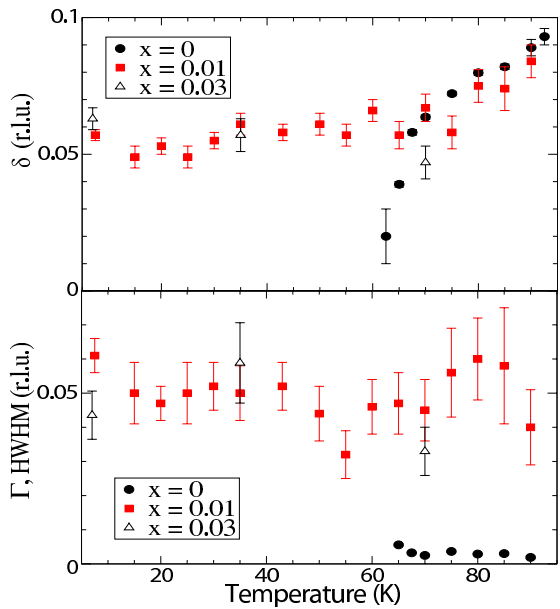}
\caption{(Color online) The temperature dependence of the incommensurate ordering wavevector, $\delta$ (top), and H-width of the incommensurate scattering, $\Gamma$ (bottom), in Ti$_{1-x}$Sc$_{x}$OCl. These parameters were determined from fitting the data in Figs. 1 and 2, as described in the text.}
\end{figure}

The temperature dependence of the incommensurate ordering wave vector, ($\delta$, 0.5, 0), is shown in the top panel of Figure 3. Between T$_{c1}$ and T$_{c2}$, the behavior of $\delta$ is remarkably similar in both pure and Sc-doped TiOCl.  In each case, the incommensurate wave vector reaches a maximum value of $\delta$ $\sim$ 0.08 near T$_{c2}$ and decreases monotonically as the temperature approaches T$_{c1}$.  At T$_{c1}$, however, $\delta$ drops rapidly to zero in TiOCl as the dimerization locks into a commensurate state, while $\delta$ remains constant at $\sim$ 0.055 down to base temperature in Ti$_{0.99}$Sc$_{0.01}$OCl and Ti$_{0.97}$Sc$_{0.03}$OCl.  This clearly illustrates that while doping strongly suppresses the transition to the commensurate dimerized state, the mechanism driving the formation of the incommensurate dimerized state remains relatively unperturbed.

The origin of the incommensurate spin-Peierls phase in TiOCl has been attributed\cite{Ruckamp} to frustrated inter-chain interactions arising from competition between in-phase and out-of-phase dimer configurations within the bilayer structure.  In this scenario, the stability of the incommensurate phase in Ti$_{0.99}$Sc$_{0.01}$OCl and Ti$_{0.97}$Sc$_{0.03}$OCl implies that the presence of non-magnetic impurities alters the balance between intra-chain and inter-chain interactions, favoring the frustrating inter-chain interactions along {\bf a}.  

The width of the incommensurate scattering, shown in the bottom panel of Figure 3, is inversely proportional to the correlation length, $\xi$.  Thus, the dramatic broadening of the incommensurate scattering observed in the H-scans of the x=0.01 and x=0.03 samples indicates very short correlation lengths along the crystallographic a-axis, $\xi$$_{a}$.  Whereas the typical correlation lengths associated with the incommensurate long-range order in TiOCl are $\sim$ 200 {\AA}, those for Ti$_{0.99}$Sc$_{0.01}$OCl and Ti$_{0.97}$Sc$_{0.03}$OCl are much shorter, on the order of $\sim$ 12 {\AA} or three unit cells.  We also note that $\xi$$_{a}$ is temperature independent in the disordered Ti$_{1-x}$Sc$_{x}$OCl, and shows little or no variation between x=0.01 and x=0.03.  

While doping is certainly expected to reduce the size of correlation lengths in the spin-Peierls state, this change in $\xi$$_{a}$ is unusual in two respects.  If one considers a simple scenario in which impurities are assumed to break dimerized chains by disrupting nearest neighbour and next-nearest neighbour interactions within the Ti-O bilayers, then x=0.01 and x=0.03 dopings should limit $\xi$$_{a}$ to approximately 76 {\AA} and 25 {\AA} respectively.  The observed correlation lengths of $\xi$$_{a}$ $\sim$ 12 {\AA} are somewhat surprising then, in that they are a factor of 2 to 6 times shorter than expected, and exhibit no discernable doping dependence.  This implies that the disruption of spin-Peierls correlation lengths in Ti$_{1-x}$Sc$_{x}$OCl is considerably stronger than our simple picture predicts, and that the effect is almost fully developed by a doping of x=0.01.

\begin{figure}
\includegraphics{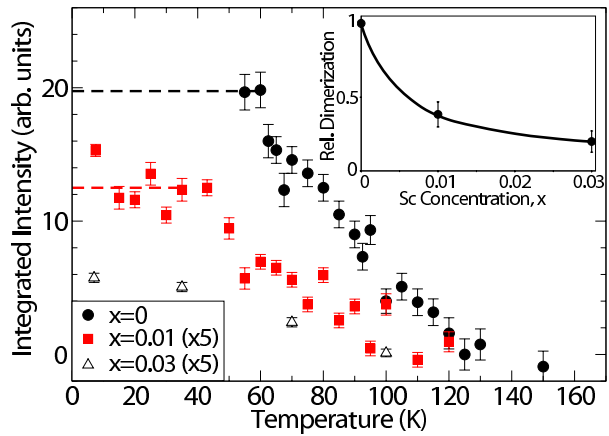}
\caption{(Color online) Temperature dependence of the integrated intensity of x-ray scattering in Ti$_{1-x}$Sc$_{x}$OCl.  All intensities were obtained by integrating over H-scans of the form (H, 1.5, 1) (as shown in Figs. 1 and 2).  The intensities of the x=0.01 and 0.03 samples have been scaled by a factor of 5. The inset shows the x-dependence of the relative spin-Peierls dimerization, determined from a comparison of the integrated scattered intensities.  All lines are intended as guides-to-the-eye. }
\end{figure}

The integrated intensity of the x-ray scattering around the commensurate (2, 1.5, 1) superlattice peak position is shown for pure and Sc-doped TiOCl in Figure 4.  These intensities come from Q-integration of H-scans through the (2, 1.5, 1) position, as shown in Figures 1 and 2.  The range of integration has been chosen to include both the commensurate and incommensurate ordering wave vectors, ensuring that all relevant dimerization scattering is captured.  In TiOCl, this integrated scattering evolves continuously from T$^*$ to T$_{c1}$, growing linearly with decreasing temperature.  Below T$_{c1}$ the integrated intensity saturates, coinciding with the development of commensurate long-range spin-Peierls order.  In the x=0.01 and x=0.03 samples, the integrated intensity begins to increase near T$_{c2}$ rather than T$^*$, and saturation occurs considerably lower than T$_{c1}$, between 40 and 50K.  The magnitude of the integrated scattering exhibits a pronounced doping dependence, as is evident from the rapid drop in observed scattering intensities that occurs with increasing x.

As the intensity of the superlattice dimerization scattering is proportional to the square of the relevant atomic displacements, a comparison of peak intensities can be used to determine the influence of impurities on the size of the spin-Peierls dimerization.  The low temperature saturation value of the integrated intensity near (2, 1.5, 1) is taken as the measure of the superlattice dimerization scattering in each of the three samples.  In order to normalize the observed scattering intensities with respect to sample volume, the weak (2, 1, 2) structural Bragg peak is used as a reference.  The structure factor for each reflection can be calculated from the expression F = $\sqrt{I sin 2 \theta}$, where I is the integrated intensity of the observed scattering and $\theta$ is the scattering angle. The relative magnitudes of the atomic displacements can then be found by comparing the ratio of F$_{obs}$(2,1.5,1) / F$_{obs}$(2,1,2) for the x=0, x=0.01 and x=0.03 samples, as shown in the inset to Figure 4.  These results imply that for impurity concentrations as low as 1 $\%$ the dimerizing displacements are less than half the size of those observed in the pure material.  A similar effect has been reported for Zn-doping in Cu$_{1-x}$Zn$_x$GeO$_{3}$\cite{Martin}, however the x-dependence in Ti$_{1-x}$Sc$_{x}$OCl is almost twice as strong.

\begin{figure}
\includegraphics{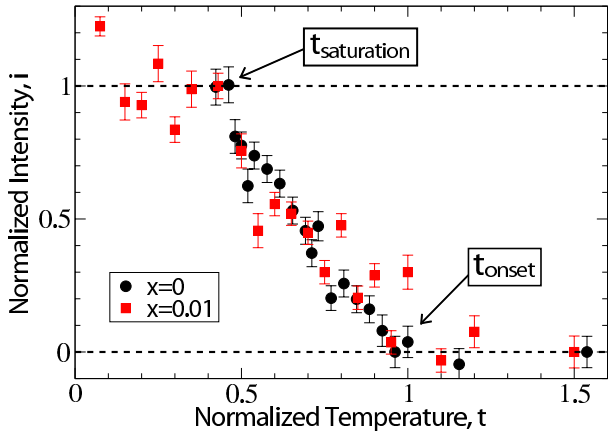}
\caption{(Color online) Normalized x-ray scattering intensity, $i = I/I_{saturation}$, as a function of normalized temperature, $t = T/T_{onset}$, in TiOCl and Ti$_{0.99}$Sc$_{0.01}$OCl.  The onset of dimerization scattering occurs at approximately T $\sim$ 130K for x = 0 and T $\sim$ 100K for x = 0.01.  All lines are intended as guides-to-the-eye. }
\end{figure}

Sc impurities are likely to perturb both the magnetic and elastic properties of TiOCl.  In addition to creating finite spin chains and uncompensated spin 1/2 moments, substitution of Sc$^{3+}$ (r=0.745 {\AA}) for Ti$^{3+}$ (r=0.67 {\AA}) will also give rise to lattice strain and local distortions.  These lattice effects may be particularly significant since it has been suggested that the frustrating inter-chain interaction in TiOCl, potentially the source of the incommensurate spin-Peierls phase, is predominantly elastic rather than magnetic in nature\cite{Schonleber}.  In doped CuGeO$_{3}$ differences in ionic radii are clearly important in disrupting spin-Peierls order.  Substituting Si$^{4+}$ (S=0, r=0.42 {\AA}) for Ge$^{4+}$ (S=0, r=0.52 {\AA}) depresses T$_{SP}$ two to three times more rapidly than doping either Zn$^{2+}$ (S=0, r=0.74 {\AA}) or Mg$^{2+}$ (S=0, r=0.72 {\AA}) onto Cu$^{2+}$ (S=1/2, r=0.72 {\AA}) sites\cite{Schoeffel, Grenier}.  Dopant size effects have also been shown to impact the critical properties of the spin-Peierls transition, even resulting in changes of universality class\cite{Lumsden}. 

Figure 5 shows the normalized x-ray scattering intensity, $i = I/I_{saturation}$, as a function of normalized temperature, $t = T/T_{onset}$, in TiOCl and Ti$_{0.99}$Sc$_{0.01}$OCl.  Here the integrated scattering intensities displayed in Figure 4 have been rescaled with respect to their low temperature saturation values, I$_{saturation}$, and the temperature ranges have been rescaled with respect to the first onset of dimerization scattering upon cooling, T$_{onset}$.  In the case of TiOCl we have taken T$_{onset}$ to be T$^*$ $\sim$ 130K, the temperature at which commensurate dimerization fluctuations initially arise, while for Ti$_{0.99}$Sc$_{0.01}$OCl we have chosen T$_{onset}$ near T$_{c2}$ at $\sim$ 100K, the temperature at which the earliest indications of incommensurate scattering appear.  Thus, the data provided in Figure 5 describes how each system evolves from the point at which dimerization (either static or fluctuating) first originates to the point at which the dimerization becomes fully developed.  Note that when scaled in this fashion, both the x=0 and x=0.01 data sets appear to follow a common linear trend, reaching saturation at roughly $t$ $\sim$ 0.45.  This correspondence is quite remarkable given that the nature of the dimerization is dramatically different in the TiOCl and Ti$_{0.99}$Sc$_{0.01}$OCl samples.

As discussed earlier in this paper, one of the most striking similarities between the measurements performed on the x=0 and x=0.01 samples is that the transition to the incommensurate spin-Peierls phase occurs at almost exactly the same temperature, near T$_{c2}$ $\sim$ 93K.  This result would seem to contradict the wealth of experimental evidence from doped CuGeO$_{3}$ which shows that the introduction of impurities leads to consistent lowering of the spin-Peierls transition temperature, T$_{SP}$\cite{Hase, Oseroff, Schoeffel, Martin, Grenier, Anderson, Mgdoped}.  However, while Sc-doping may not alter T$_{c2}$, it does produce a substantial drop in the onset temperature and saturation temperature of the spin-Peierls dimerization. Between x=0 and x=0.01 the values of T$_{onset}$ and T$_{saturation}$ decrease by approximately 25 \%, an effect very similar in size to the suppression of T$_{SP}$ reported for doped CuGeO$_{3}$ (typically between 15-45 \% of T$_{SP}$ for every 1 \% of dopants added to the system).  These observations suggest that the most fundamental temperature scale for the spin-Peierls behavior in TiOCl and Ti$_{1-x}$Sc$_{x}$OCl is not defined by the phase transitions at T$_{c1}$ and T$_{c2}$, but rather by the appearance and development of the dimerizing lattice displacement.


\section{Conclusions}

In conclusion, we have shown that quenched non-magnetic Sc impurities suppress commensurate spin-Peierls order and fluctuations in Ti$_{1-x}$Sc$_{x}$OCl, even at the x=0.01 level.  A short-range ordered, incommensurate state arises near T$_{c2}$ of the pure material, and this state is maintained down to the lowest temperatures measured ($\sim$ 7K).  The stability of the incommensurate spin-Peierls state, together with the absence of phase coexistence between T$_{c1}$ and T$_{c2}$, suggests that Sc-doping acts to effectively enhance the strength of frustrating inter-chain interactions within the Ti-O bilayers.  The presence of Sc impurities also serves to break up one-dimensional dimer chains, dramatically decreasing the characteristic correlation lengths of the spin-Peierls state. In addition, Sc-doping is found to severely affect the properties of the dimerizing atomic displacements which are associated with the formation of the spin-Peierls state, reducing both the average size of the dimerization and the temperature at which it first occurs. We hope that this work may guide and inform future theoretical and experimental studies of the doped titanium oxyhalide systems.   

\begin{acknowledgments}

The authors would like to acknowledge helpful discussions with T. Imai, A. Aczel, and J. Ruff.  This work was supported by NSERC of Canada and NSC of Taiwan.

\end{acknowledgments}

%
%
%
%
%
%
%
%
%
%

\end{document}